\documentstyle[12pt]{article}
\newcommand{\bd}{\begin{document}}
\newcommand{\ed}{\end{document}}
\newcommand{\bc}{\begin{center}}
\newcommand{\ec}{\end{center}}
\newcommand{\be}{\begin{eqnarray}}
\newcommand{\ee}{\end{eqnarray}}
\renewcommand{\thefootnote}{\alph{footnote}}
\newcommand{\se}{\section}
\newcommand{\sse}{\subsection}
\newcommand{\bi}{\bibitem}
\def\figcap{\section*{Figure Captions\markboth
     {FIGURECAPTIONS}{FIGURECAPTIONS}}\list
     {Figure \arabic{enumi}:\hfill}{\settowidth\labelwidth{Figure 999:}
     \leftmargin\labelwidth
     \advance\leftmargin\labelsep\usecounter{enumi}}}
\let\endfigcap\endlist \relax

\begin{document}

%\begin{center}
\begin{titlepage}

 \vskip 0.5in
 \null
\begin{center}
 \vspace{.15in}
{\LARGE {\bf
Study of $\Lambda _{b}\rightarrow \Lambda \ \nu \ \bar{\nu}$ with
Polarized Baryons}
}\\
\vspace{1.0cm}  \par
 \vskip 2.1em
 {\large
  \begin{tabular}[t]{c}
{\bf Chuan-Hung Chen$^a$ and C.~Q.~Geng$^b$}
\\
\\
       {\sl ${}^a$Department of Physics, National Cheng Kung University}
\\   {\sl  $\ $Tainan, Taiwan,  Republic of China }
\\
\\
{\sl ${}^b$Department of Physics, National Tsing Hua University}
\\  {\sl  $\ $ Hsinchu, Taiwan, Republic of China }
\\
   \end{tabular}}
 \par \vskip 5.3em

%\title{Study of $\Lambda _{b}\rightarrow \Lambda \ \nu \ \bar{\nu}$ with
%Polarized Baryons}
%\author{}
%\maketitle

 {\Large\bf Abstract}
\end{center}
%\begin{abstract}
We investigate the decay of $\Lambda _{b}\rightarrow \Lambda \ \nu \ \bar{\nu}$
with the polarized baryons of $\Lambda _{b}$ and $\Lambda $.
With the most general hadronic form
factors, we first study the decay branching ratio and then derive
the longitudinal, normal and transverse polarizations of $\Lambda$ in
terms of the spin unit vectors of $\Lambda _{b}$ and $\Lambda $
 and the momentum of $\Lambda $. A polarization of $\Lambda _{b}$ is also
discussed.

%\end{abstract}

\end{titlepage}

\section{Introduction}

Recently, some of the interest in flavor physics has been focused in the
rare decays related to $b\to s l\;\bar{l}$ induced by the flavor changing
neutral current (FCNC) due to the CLEO measurement of the radiative $b \to
s\gamma$ decay \cite{cleo}. In the standard model, these rare decays occur
at loop level and provide us with information on the parameters of the
Cabibbo-Kobayashi-Maskawa (CKM) matrix elements \cite{ckm} as well as
various hadronic form factors. The corresponding rare decays of heavy
hadrons such as $\Lambda_b\to\Lambda l^+l^-$ have been studied in the
literature \cite{Lb1,Lb2}.

In this paper, we investigate the decay of $\Lambda _{b}\rightarrow \Lambda
\ \nu \ \bar{\nu}$ with the polarized baryons of $\Lambda _{b}$ and $\Lambda
$. To study the decay, we shall use the most general hadronic form factors
for the $\Lambda_b \to \Lambda$ transition. It is clear that this decay is
free of long distance uncertainty like other di-neutrino decays of mesons
\cite{LD}. However, there are many form factors when one evaluates the
hadronic matrix elements between $\Lambda_b$ and $\Lambda$, which are hard
to be calculated since they are related to the non-perturbative effect of
QCD. It is known that for heavy particle decays, the heavy quark effective
theory (HQET) could reduce the number of form factors and supply the
information with respect to their relative size. In our numerical
calculations, we shall consider the cases with and without HQET. In our
discussions of the decay branching ratio and the polarizations of $\Lambda$
and $\Lambda _{b}$, we shall study the standard model and new physics such
as those with right-handed hadronic currents.

The paper is organized as follows. In Sec.~2, we study the effective
Hamiltonian for the di-neutrino decay of $\Lambda_b \to \Lambda \nu \bar{\nu}
$ and form factors in the $\Lambda_b \to \Lambda$ transition. In Sec.~3, we
derive the general forms of the differential decay rate and the
longitudinal, normal and transverse polarizations of $\Lambda$. A
polarization of $\Lambda _{b}$ is also discussed. In Sec.~4, we give the
numerical analysis. We present our conclusions in Sec.~5.

\section{Effective Hamiltonian and Form factors}

We start by writing the effective Hamiltonian for the inclusive process of $%
b\rightarrow s\nu \bar{\nu}$ as
\begin{equation}
{\cal H}\left( b\rightarrow s\ \nu \ \bar{\nu}\right) =\left[ C_{L}\ \bar{s}%
\ \gamma _{\mu }\ P_{L}b+C_{R}\ \bar{s}\ \gamma _{\mu }\ P_{R}b\right] \bar{%
\nu}\ \gamma ^{\mu }P_{L}\nu \,,  \label{Heff1}
\end{equation}
with $P_{L\left( R\right) }=\left( 1\mp \gamma _{5}\right) /2$, where we
have assumed that the theories contain only $V-A$ and $V\pm A$-type
interactions for the lepton and quark sectors, respectively.
This can be justified if there is no contribution to the decay from
right-handed neutrinos.
Moreover, since the neutrino masses are very small, we expect that
in our study only
 $V-A$-type of interactions for the lepton sector is important.
In Eq. (\ref
{Heff1}), $C_{L,R}$  are defined by
\begin{eqnarray}
C_{L} &=&C^{SM}\left( 1+\delta _{L}\right) \,,  \nonumber \\
C_{R} &=&C^{SM}\delta _{R}\,,  \nonumber \\
C^{SM} &=&{\frac{G_{F}}{\sqrt{2}}}{\frac{\alpha _{em}}{\pi }}{\frac{2}{\sin
^{2}\theta _{w}}}V_{tb}V_{ts}^{*}X\left( x_{t}\right) \,,
\end{eqnarray}
with
\begin{equation}
X\left( x_{t}\right) =\eta _{QCD}\frac{x_{t}}{8}\left[ \frac{x_{t}+2}{x_{t}-1%
}+\frac{3x_{t}-6}{\left( x_{t}-1\right) ^{2}}\ln x_{t}\right] ,
\end{equation}
where
$C^{SM}$ stands for the contribution from the standard
model, $\delta _{L,R}$ denote the effects from new physics,
$x_t=m_t^2/M_W^2$, and $\eta
_{QCD}=0.985$ is the QCD correction \cite{Buchalla}. In the standard model,
one has that $\delta _{L,R}=0$.
The constraints on  $\delta _{L,R}$ will be discussed in Sec. 4.
For calculating exclusive decays such as $%
\Lambda _{b}\rightarrow \Lambda \ \nu \ \bar{\nu}$, one has to evaluate the
hadronic matrix element of the $\Lambda _{b}\to \Lambda $ transition. In
general, one can express the vector and axial vector currents for the
transition as
\begin{eqnarray}
\left\langle \Lambda \right| \bar{s}\ \gamma _{\mu }\ b\left| \Lambda
_{b}\right\rangle  &=&f_{1}\bar{u}_{\Lambda }\gamma _{\mu }u_{\Lambda
_{b}}+if_{2}\bar{u}_{\Lambda }\sigma _{\mu \nu }\ q^{\nu }u_{\Lambda
_{b}}+f_{3}q_{\mu }\bar{u}_{\Lambda }u_{\Lambda _{b}},  \nonumber \\
\left\langle \Lambda \right| \bar{s}\ \gamma _{\mu }\gamma _{5}\ b\left|
\Lambda _{b}\right\rangle  &=&g_{1}\bar{u}_{\Lambda }\gamma _{\mu }\gamma
_{5}u_{\Lambda _{b}}+ig_{2}\bar{u}_{\Lambda }\sigma _{\mu \nu }\ q^{\nu
}\gamma _{5}u_{\Lambda _{b}}+g_{3}q_{\mu }\bar{u}_{\Lambda }\gamma
_{5}u_{\Lambda _{b}}\,,  \label{FF}
\end{eqnarray}
where $f_{i}$ and $g_{i}$ ($i=1,2,3$) are the form factors and $q$ is the
momentum difference of the baryons, $i.e.$,
$q=p_{\Lambda _{b}}-p_{\Lambda }$.
It is clear that the terms corresponding to $f_{3}$ and $g_{3}$ have no
contribution to the di-neutrinos decays for the massless neutrino case
in the standard model,
while that for the massive neutrinos, the effect is negligible.
Therefore, we shall consider only
the four independent form factors of $f_{1,2}$ and $g_{1,2}$ in Eq. (\ref{FF}%
). From Eqs. (\ref{Heff1}) and (\ref{FF}), we obtain the general form of the
transition matrix element for $\Lambda _{b}\rightarrow \Lambda \nu \bar{\nu}$
as
\be
{\cal M}=(F_{1}\bar{u}_{\Lambda }\gamma _{\mu }u_{\Lambda _{b}}+G_{1}\bar{u}%
_{\Lambda }\gamma _{\mu }\gamma _{5}u_{\Lambda _{b}}+F_{2}\bar{u}_{\Lambda }%
\not{q}\gamma _{\mu }u_{\Lambda _{b}}+G_{2}\bar{u}_{\Lambda }\not{q}\gamma
_{\mu }\gamma _{5}u_{\Lambda _{b}}\ )\bar{\nu}\ \gamma ^{\mu }P_{L}\nu ,
\label{matrix}
\ee
where
\be
F_{i}=\frac{C_{R}+C_{L}}{2}f_{i}\,,\quad G_{i}=\frac{C_{R}-C_{L}}{2}%
g_{i}\,,\quad (i=1,2)\,.
\ee

In HQET, the matrix elements in Eq. (\ref{FF}) can be simplified.
Explicitly, from Ref. \cite{MR}, one has that
\begin{equation}
\left\langle \Lambda \right| \bar{s}\ \Gamma \ b\left| \Lambda
_{b}\right\rangle =\bar{u}_{\Lambda }\left( {\cal F}_{1}+\not{v}{\cal F}%
_{2}\right) \Gamma u_{\Lambda _{b}}\,,  \label{HFF}
\end{equation}
where $\Gamma $ denotes the possible Dirac matrix and $v=P_{\Lambda
_{b}}/M_{\Lambda _{b}}$ is the four-velocity of $\Lambda _{b}$. Comparing
Eq. (\ref{HFF}) with Eq. (\ref{FF}), we get
\begin{eqnarray}
f_{1} &=&g_{1}\;=\;{\cal F}_{1}+\sqrt{r}{\cal F}_{2},  \nonumber \\
f_{2} &=&g_{2}\;=\;\frac{1}{M_{\Lambda _{b}}}{\cal F}_{2},  \label{hqet}
\end{eqnarray}
where $r=M_{\Lambda }^{2}/M_{\Lambda _{b}}^{2}$. Clearly, based on HQET, the
form factors corresponding to the vector current are the same as that to the
axial vector one, and the parts of the electric and magnetic moment are
suppressed by the mass of the heavy particle. Therefore, there are only two
independent form factors ${\cal F}_{1,2}$ and the form factors of the
hadronic vector and axial vector currents are larger than that of the
hadronic electric and magnetic currents in HQET.

\section{Differential decay rate and Polarizations}

To study the polarized baryon $(B=\Lambda _{b}$ or $\Lambda)$, we write the
four-spin vector of the baryon as
\begin{equation}
s_{B}^{0}=\frac{\vec{p}_{B}\cdot \hat{\xi}_{B}}{M_{B}},\qquad \vec{s}_{B}=%
\hat{\xi}_{B}+\frac{s_{B}^{0}}{E_{B}+M_{B}}\vec{p}_{B}\,,
\end{equation}
where $p_B$ is the momentum of $B$ and $\hat{\xi}_{B}$ is the unit vector
along the baryon spin in its rest frame.

In the $\Lambda_{b}$ rest frame, we choose the unit vectors along the
longitudinal, normal, transverse components of the $\Lambda $ polarization
as $\hat{e}_i\ (i=L,N,T)$, defined by
\begin{eqnarray}
\hat{e}_{L} &=&\frac{\vec{p}_{\Lambda }}{\left| \vec{p}_{\Lambda }\right| }%
\,,  \nonumber \\
\hat{e}_{N} &=&\hat{e}_{L}\times \left( \hat{\xi}_{\Lambda _{b}}\times \hat{e%
}_{L}\right) \,,  \nonumber \\
\hat{e}_{T} &=&\hat{\xi}_{\Lambda _{b}}\times \hat{e}_{L} \,,  \label{uv}
\end{eqnarray}
respectively, where $\vec{p}_{\Lambda }$ is the momentum of $\Lambda$.

The partial decay rate for $\Lambda _{b}(p_{\Lambda_{b}},s_{\Lambda_{b}})
\to \Lambda (p_{\Lambda},s_{\Lambda}) \nu (p_{1})\bar{\nu} (p_{2})$ is given
by
\begin{eqnarray}
d\Gamma &=&\frac{1}{4M_{\Lambda _{b}}}{\cal MM}^{\dagger }d\Phi\,,  \nonumber
\\
d\Phi &=&\left( 2\pi \right) ^{4}\delta ^{4}\left( p_{\Lambda
_{b}}-p_{\Lambda }-p_{1}-p_{2}\right) \frac{d^{3}p_{\Lambda }}{\left( 2\pi
\right) ^{3}2E_{\Lambda }}\frac{d^{3}p_{1}}{\left( 2\pi \right) ^{3}2E_{1}}%
\frac{d^{3}p_{2}}{\left( 2\pi \right) ^{3}2E_{2}}\,.  \label{partial}
\end{eqnarray}
In the $\Lambda_{b}$ rest frame, by integrating the phase space of $\nu $
and $\bar{\nu}$, from Eqs. (\ref{matrix}) and (\ref{partial}) the partial
decay rate in terms of the energy and polarizations of $\Lambda $ is then
given by
\begin{eqnarray}
d\Gamma &=&\frac{1}{4}\left[ 1+\frac{I_{2}}{I_{1}}\hat{e}_{L}\cdot \hat{\xi}%
_{\Lambda _{b}}\right] \left[ 1+\vec{P}_{\Lambda }\cdot \hat{\xi}_{\Lambda
}\right] d\Gamma ^{0}\,,  \label{diffrate1}
\end{eqnarray}
with
\begin{eqnarray}
d\Gamma ^{0} &=&3\frac{G_{F}^{2}\alpha _{em}^{2}\left|
V_{tb}V_{ts}^{*}\right| ^{2}}{384\pi ^{6}M_{\Lambda _{b}}}\sqrt{E_{\Lambda
}^{2}-M_{\Lambda }^{2}}\ I_{1}\ dE_{\Lambda }\ d\Omega _{\Lambda }\,
\label{diffrate2}
\end{eqnarray}
and
\begin{equation}
\vec{P}_{\Lambda }=\frac{1}{1+\frac{I_{2}}{I_{1}}\hat{e}_{L}\cdot \hat{\xi}%
_{\Lambda _{b}}}\left[ \left( \frac{I_{3}}{I_{1}}+\frac{I_{4}}{I_{1}}\hat{e}%
_{L}\cdot \hat{\xi}_{\Lambda _{b}}\right) \hat{e}_{L}+\frac{I_{5}}{I_{1}}%
\hat{e}_{N}+\frac{I_{6}}{I_{1}}\hat{e}_{T}\right]\,,  \label{pola}
\end{equation}
where the factor 3 in Eq. (\ref{diffrate2}) represents three families of
neutrinos and $I_i$ are defined by
\begin{eqnarray}
I_{1} &=&\left( \left| F_{1}\right| ^{2}+\left| G_{1}\right| ^{2}\right)
\left( q^{2}p_{\Lambda _{b}}\cdot p_{\Lambda }+2p_{\Lambda _{b}}\cdot
qp_{\Lambda }\cdot q\right) +3\left( \left| F_{1}\right| ^{2}-\left|
G_{1}\right| ^{2}\right) M_{\Lambda _{b}}M_{\Lambda }q^{2}  \nonumber \\
&&-\left( \left| F_{2}\right| ^{2}+\left| G_{2}\right| ^{2}\right) \left(
q^{4}p_{\Lambda _{b}}\cdot p_{\Lambda }-4q^{2}p_{\Lambda _{b}}\cdot
qp_{\Lambda }\cdot q\right) -3\left( \left| F_{2}\right| ^{2}-\left|
G_{2}\right| ^{2}\right) M_{\Lambda _{b}}M_{\Lambda }q^{4}  \nonumber \\
&&-6M_{\Lambda _{b}}q^{2}p_{\Lambda }\cdot q\left( \mathop{\rm Re}%
F_{2}F_{1}^{*}-\mathop{\rm Re}G_{2}G_{1}^{*}\right) +6M_{\Lambda
}q^{2}p_{\Lambda _{b}}\cdot q\left( \mathop{\rm Re}F_{2}F_{1}^{*}+%
\mathop{\rm Re}G_{2}G_{1}^{*}\right) \,,  \label{I1} \\
I_{2} &=&-2\mathop{\rm Re}F_{1}G_{1}^{*}M_{\Lambda _{b}}\left(
q^{2}-2p_{\Lambda }\cdot q\right) \sqrt{E_{\Lambda }^{2}-M_{\Lambda }^{2}}%
+2M_{\Lambda _{b}}q^{2}\sqrt{E_{\Lambda }^{2}-M_{\Lambda }^{2}}
\times \nonumber \\
&& \left[ \mathop{\rm Re}F_{2}G_{1}^{*}\left( M_{\Lambda
_{b}}+3M_{\Lambda }\right) -\mathop{\rm Re}F_{1}G_{2}^{*}\left( M_{\Lambda
_{b}}-3M_{\Lambda }\right) +\mathop{\rm Re}F_{2}G_{2}^{*}\left(
q^{2}+4p_{\Lambda }\cdot q\right) \right] \,,  \label{I2} \\
I_{3} &=&2\mathop{\rm Re}F_{1}G_{1}^{*}M_{\Lambda _{b}}\left(
q^{2}+2p_{\Lambda _{b}}\cdot q\right) \sqrt{E_{\Lambda }^{2}-M_{\Lambda }^{2}%
}-2M_{\Lambda _{b}}q^{2}\sqrt{E_{\Lambda }^{2}-M_{\Lambda }^{2}}
\times \nonumber
\\
&& \left[ \mathop{\rm Re}F_{2}G_{1}^{*}\left( M_{\Lambda }+3M_{\Lambda
_{b}}\right) +\mathop{\rm Re}F_{1}G_{2}^{*}\left( M_{\Lambda }-3M_{\Lambda
_{b}}\right) -\mathop{\rm Re}F_{2}G_{2}^{*}\left( q^{2}-4p_{\Lambda
_{b}}\cdot q\right) \right] \,,  \label{I3} \\
I_{4} &=&\left( \left| F_{1}\right| ^{2}-\left| G_{1}\right| ^{2}\right)
\left[ \frac{M_{\Lambda _{b}}}{M_{\Lambda }}\left( E_{\Lambda
}^{2}-M_{\Lambda }^{2}\right) \left( 2M_{\Lambda }^{2}-q^{2}-2p_{\Lambda
}\cdot q\right) \right.  \nonumber \\
&&\left. -\frac{E_{\Lambda }}{M_{\Lambda }}\left( 2p_{\Lambda _{b}}\cdot q\
p_{\Lambda }\cdot q-q^{2}P_{\Lambda _{b}}\cdot p_{\Lambda }\right) \right]
+\left( \left| F_{2}\right| ^{2}+\left| G_{2}\right| ^{2}\right)  \nonumber
\\
&&\times \left[ -M_{\Lambda _{b}}E_{\Lambda }q^{4}+\frac{M_{\Lambda _{b}}}{%
M_{\Lambda }}q^{4}\left( E_{\Lambda }^{2}-M_{\Lambda }^{2}\right) \right]
+\left( \left| F_{2}\right| ^{2}-\left| G_{2}\right| ^{2}\right)  \nonumber
\\
&&\times \left[ \frac{E_{\Lambda }}{M_{\Lambda }}p_{\Lambda _{b}}\cdot
p_{\Lambda }q^{4}+4M_{\Lambda _{b}}^{2}\left( E_{\Lambda }^{2}-M_{\Lambda
}^{2}\right) q^{2}\right] +2\left( \mathop{\rm Re}F_{2}F_{1}^{*}+\mathop{\rm
Re}G_{2}G_{1}^{*}\right)  \nonumber \\
&&\times \left[ -\frac{M_{\Lambda _{b}}}{M_{\Lambda }}E_{\Lambda
}q^{2}p_{\Lambda }\cdot q+\frac{M_{\Lambda _{b}}^{2}}{M_{\Lambda }}%
q^{2}\left( E_{\Lambda }^{2}-M_{\Lambda }^{2}\right) \right] +2\left( %
\mathop{\rm Re}F_{2}F_{1}^{*}-\mathop{\rm Re}G_{2}G_{1}^{*}\right)  \nonumber
\\
&&\times \left[ E_{\Lambda }q^{2}p_{\Lambda }\cdot q+M_{\Lambda
_{b}}q^{2}\left( E_{\Lambda }^{2}-M_{\Lambda }^{2}\right) \right] \,,
\label{I4} \\
I_{5} &=&-\left( \left| F_{1}\right| ^{2}-\left| G_{1}\right| ^{2}\right)
\left( 2p_{\Lambda _{b}}\cdot q\ p_{\Lambda }\cdot q-q^{2}P_{\Lambda
_{b}}\cdot p_{\Lambda }\right) +\left( \left| F_{2}\right| ^{2}-\left|
G_{2}\right| ^{2}\right) q^{4}P_{\Lambda _{b}}\cdot p_{\Lambda }  \nonumber
\\
&&-\left( \left| F_{2}\right| ^{2}+\left| G_{2}\right| ^{2}\right)
M_{\Lambda _{b}}M_{\Lambda }q^{4}+2\left( \mathop{\rm Re}F_{2}F_{1}^{*}-%
\mathop{\rm Re}G_{2}G_{1}^{*}\right) M_{\Lambda }q^{2}p_{\Lambda _{b}}\cdot q
\nonumber \\
&&-2\left( \mathop{\rm Re}F_{2}F_{1}^{*}+\mathop{\rm Re}G_{2}G_{1}^{*}%
\right) M_{\Lambda _{b}}q^{2}p_{\Lambda }\cdot q\,,  \label{I5} \\
I_{6} &=&2\mathop{\rm Im}F_{1}G_{1}^{*}M_{\Lambda _{b}}\sqrt{E_{\Lambda
}^{2}-M_{\Lambda }^{2}}\left( q^{2}-2P_{\Lambda _{b}}\cdot q\right)
+2M_{\Lambda _{b}}q^{2}\sqrt{E_{\Lambda }^{2}-M_{\Lambda }^{2}}  \nonumber \\
&&\times \left[ M_{\Lambda _{b}}\left( \mathop{\rm Im}F_{2}G_{1}^{*}-%
\mathop{\rm Im}F_{1}G_{2}^{*}\right) -M_{\Lambda }\left( \mathop{\rm Im}%
F_{2}G_{1}^{*}+\mathop{\rm Im}F_{1}G_{2}^{*}\right) +q^{2}\mathop{\rm Im}%
F_{2}G_{2}^{*}\right] \,.  \label{I6}
\end{eqnarray}
%We note that the small neutrino mass effects to the rate
%are also negligible.
Here the kinematics and the relationships for the form factors are given as
follows:
\begin{eqnarray}
q^{2} &=& M_{\Lambda _{b}}^{2}+M_{\Lambda }^{2}-2M_{\Lambda _{b}}E_{\Lambda
}\,,  \nonumber \\
p_{\Lambda _{b}}\cdot p_{\Lambda } &=&M_{\Lambda _{b}}E_{\Lambda }\,,
\nonumber \\
p_{\Lambda _{b}}\cdot q &=&M_{\Lambda _{b}}^{2}-M_{\Lambda _{b}}E_{\Lambda
}\,,  \nonumber \\
p_{\Lambda }\cdot q &=&M_{\Lambda _{b}}E_{\Lambda }-M_{\Lambda }^{2}\,,
\end{eqnarray}
and
\begin{eqnarray}
F_{j}F_{k}^{*} &=&f_{j}f_{k}\left[ \frac{\left| C_{R}\right| ^{2}+\left|
C_{L}\right| ^{2}}{4}+\frac{1}{2}\mathop{\rm Re}C_{L}C_{R}^{*}\right] \,,
\nonumber \\
G_{j}G_{k}^{*} &=&g_{j}g_{k}\left[ \frac{\left| C_{R}\right| ^{2}+\left|
C_{L}\right| ^{2}}{4}-\frac{1}{2}\mathop{\rm Re}C_{L}C_{R}^{*}\right] \,,
\nonumber \\
F_{j}G_{k}^{*} &=&f_{j}g_{k}\left[ \frac{\left| C_{R}\right| ^{2}-\left|
C_{L}\right| ^{2}}{4}+\frac{i}{2}\mathop{\rm Im}C_{L}C_{R}^{*}\right] \,,
\nonumber \\
\left| F_{i}\right| ^{2}+\left| G_{i}\right| ^{2} &=&\frac{1}{4}\left(
f_{i}^{2}+g_{i}^{2}\right) \left( \left| C_{R}\right| ^{2}+\left|
C_{L}\right| ^{2}\right) +\frac{1}{2}\left( f_{i}^{2}-g_{i}^{2}\right) %
\mathop{\rm Re}C_{L}C_{R}^{*}\,,  \nonumber \\
\left| F_{i}\right| ^{2}-\left| G_{i}\right| ^{2} &=&\frac{1}{4}\left(
f_{i}^{2}-g_{i}^{2}\right) \left( \left| C_{R}\right| ^{2}+\left|
C_{L}\right| ^{2}\right) +\frac{1}{2}\left( f_{i}^{2}+g_{i}^{2}\right) %
\mathop{\rm Re}C_{L}C_{R}^{*}\,.  \label{ffgg}
\end{eqnarray}

  From Eq. (\ref{diffrate2}), we can find the decay rate of $\Lambda_b\to
\Lambda \nu\bar{\nu}$ by integrating the energy of $\Lambda $. The solid
angle and the numerical values for the decay branching ratio are shown in
the next section. In the standard model, the dominant and subdominant
contributions to decay rate in Eq. (\ref{diffrate2}) are the first and last
terms in Eq. (\ref{I1}), which are proportional to $\left(
f_{1}^{2}+g_{1}^{2}\right) $ and $\left(f_{1}f_{2}+g_{1}g_{2}\right) $,
respectively. Since the form factors of $f_{2}$ and $g_{2}$ are
negative \cite{Lb1},
the term relating to $\left(f_{1}f_{2}+g_{1}g_{2}\right) $ gives destructive
contribution to the decay rate.

The three components of $\vec{P}_{\Lambda}$ in Eq. (\ref{pola}),
corresponding to the longitudinal, normal and transverse polarization
asymmetries of $\Lambda$, can be also defined by
\begin{equation}
P_{i}=\frac{d\Gamma \left( \hat{\xi}_{B}\cdot \hat{e}_{i}=1\right) -d\Gamma
\left( \hat{\xi}_{B}\cdot \hat{e}_{i}=-1\right) }{d\Gamma \left( \hat{\xi}%
_{B}\cdot \hat{e}_{i}=1\right) +d\Gamma \left( \hat{\xi}_{B}\cdot \hat{e}%
_{i}=-1\right) }\,,\quad (i=L,N,T)\,,  \label{pasy}
\end{equation}
respectively.

When $\Lambda $ is not polarized, that is $\hat{\xi}_{\Lambda }=0$, from Eq.
(\ref{diffrate1}) by summing the spin of $\Lambda$ we obtain
\begin{eqnarray}
d\Gamma &=& {\frac{d\Gamma^0 }{2}} \left( 1+\alpha _{\Lambda _{b}}\hat{\xi}%
_{\Lambda _{b}}\cdot \hat{e}_{L}\right)  \label{alpha1}
\end{eqnarray}
with
\begin{eqnarray}
\alpha _{\Lambda _{b}} &=&\frac{I_{2}}{I_{1}}\,.
\end{eqnarray}
 From the above equation, we may write the polarization of $\Lambda _{b}$
as $P_{\Lambda_b}$ defined by
\begin{eqnarray}
P_{\Lambda_b} &\equiv & \alpha _{\Lambda_b}\,,  \label{plb}
\end{eqnarray}
when $\hat{\xi}_{\Lambda }=0$. For unpolarized $\Lambda_{b}$, $i.e.$, $\hat{%
\xi}_{\Lambda _{b}}=0$, one obtains that
\begin{eqnarray}
\vec{P}_{\Lambda} &=& \alpha _{\Lambda }\hat{e}_{L}\,,  \label{alpha2}
\end{eqnarray}
where
\begin{eqnarray}
\alpha _{\Lambda } &=& \frac{I_{3}}{I_{1}}\,,
\end{eqnarray}
which implies that the $\Lambda $ polarization is purely longitudinal. In
this case, one has that $P_L=\alpha _{\Lambda }$ and $P_N=P_T=0$. We note
that, in the standard model, the longitudinal polarization of $\Lambda$ in
Eq. (\ref{alpha2}) and the polarization of $\Lambda_b$ in Eq. (\ref{plb})
are independent of the couplings due to the cancellations between $I_{3,2}$
and $I_1$, respectively. Thus, these polarizations in $\Lambda
_{b}\rightarrow \Lambda \nu \bar{\nu}$ are constants and the hadronic form
factors are the only theoretical uncertainties.

We note that the transverse component ($P_T$) of the $\Lambda$ polarization
in Eq. (\ref{pola}) is a T-odd quantity. A nonzero value of $P_T$ could
indicate CP violation. In the standard model, since there is no CP violating
phase in the CKM elements of $V_{tb}V_{ts}^{*}$, it cannot induce $P_T$ in
the decay of $\Lambda _{b}\rightarrow \Lambda \nu \bar{\nu}$ with polarized
initial and final baryons. Clearly, if the transverse $\Lambda$ polarization
is measured in an experiment, it could tell us that there exist new CP
violating sources and new types of interactions as well in nature.

\section{Numerical Analysis}

In this section, we study the numerical values of the decay branching ratio
and polarizations of $\Lambda_{b}\rightarrow \Lambda \nu \bar{\nu}$ in the
standard model and theories of new physics, respectively.

As mentioned in Sec. I, in general there are four hadronic form factors, $%
f_{i}$ and $g_{i}\ (i=1,2)$, for the $\Lambda _{b}\rightarrow \Lambda $
transition. But, under the assumption of HQET, the four, related to ${\cal F}%
_{1}$ and ${\cal F}_{2}$, becomes two. For simplicity, we take HQET as a
good approximation and use the results of Ref. \cite{Lb1} where ${\cal F}%
_{1} $ and ${\cal F}_{2}$ were calculated by using the QCD sum rule
approach. However, in the approach there is a undetermined parameter, so
called the Borel parameter $(M)$, introduced for the suppression of the
contribution from the higher excited and continuum states. It is found that $%
1.5\ GeV\leq M\leq 1.9\ GeV$ from the analysis of Ref. \cite{Lb1}. For our
numerical calculations, if it is not mentioned further, we take $M=1.7$ as
an input value.

The new physics parameters of $\delta _{L}$ and $\delta _{R}$
can be limited by the decay branching ratio of
$B\rightarrow X_{s}\nu \bar{\nu}$,
given by \cite{Buras}
%\be
%\Gamma =3\frac{m_{b}^{5}}{768\pi ^{5}}\left( \frac{G_{F}\alpha
%_{em}V_{ts}^{*}V_{tb}}{\sin ^{2}\theta _{w}}X\left( x_{t}\right) \right)
%^{2}\left( \left( 1+\delta _{L}\right) ^{2}+\delta _{R}^{2}\right)\,.
\be
\frac{B\left( B\rightarrow X_{s}\nu \bar{\nu}\right) }{B\left( B\rightarrow
X_{c}e\bar{\nu}\right) }=\frac{3\alpha _{em}^{2}}{4\pi \sin ^{4}\theta _{w}}%
\frac{\left| V_{ts}\right| ^{2}}{\left| V_{cb}\right| ^{2}}\frac{X^{2}\left(
x_{t}\right) }{f\left( z\right) }\frac{\eta }{\kappa \left( z\right) }\left(
| 1+\delta _{L}| ^{2}+|\delta _{R}|^{2}\right)
\label{Const1}
\ee
where $f(z) $ is the phase-space factor,
 $\kappa (z)$
is the QCD correction for $B\to X_{c}e\bar{\nu}$,
$z=m_c/m_b$,
and $\eta $ denotes the QCD correction to the matrix element of
$b\to s\nu \bar{\nu}$.
By taking $B( B\to X_{c}e\bar{\nu}) =11\%$, $f(z) =0.49$,
$\kappa ( z) =0.88$, $\eta =0.83$ and
$m_t(m_t) =165\ GeV$, and using the limit of
$B( B\to X_{s}\nu
\bar{\nu}) <7.7\times 10^{-4}$ \cite{ALEPH},
we obtain the constraint on $\delta_{L,R}$ as follows:
%\be
%B\left( B\rightarrow X_{s}\nu \bar{\nu}\right) =4.0\times 10^{-5}\left(
%\left( 1+\delta _{L}\right) ^{2}+\delta _{R}^{2}\right) \,.
%\label{bnunu}
%\ee
%Based on the data of ALEPH Collaboration \cite{ALEPH}, the limit on $%
%B\rightarrow X_{s}\nu \bar{\nu}$ is given by
%$B\left( B\rightarrow X_{s}\nu
\be
| 1+\delta_L|^2+|\delta_R|^2<19.3\,.
\label{Const2}
\ee
Clearly, as we can see from Eq. (\ref{Const2}), large ranges for
the values of $\delta_{L,R}$ are allowed.
However, it will be shown in Sec. 4.2 that  by requring
the longitudinal polarization of $\Lambda$ being less than one,
the parameters of $\delta_{L,R}$ can be further constrained.

\subsection{Decay branching ratio}

\noindent $\bullet$ {\em In the standard model}

In this subsection, we estimate the decay branching ratio of $%
\Lambda_{b}\rightarrow \Lambda \nu \bar{\nu}$ in the standard model with and
without the assumption of HQET. It is known that so far there is no full
calculation on the form factors of vector and axial vector currents for
baryonic decays. Therefore, for the non-HQET case, we still use the results
of Ref. \cite{Lb1} but take several values of $g_{i}/f_{i}\ ( i=1,2) $
around one required by HQET. In Table 1, we show the decay branching ratio
with different ratios of $g_i/f_{i}$. From the table we clearly see that
even the differences of $g_{i}/f_{i}$ are up to $20\%$, the influence on the
decay branching ratio is only at a few percent level. Since we know that $%
f_{2}\ ( g_{2}) \ll f_{1}\ ( g_{1}) $ by Eq. (\ref{hqet}) due to the
suppression of the heavy quark mass, if one excludes the contributions of $%
f_{2}$ and $g_{2}$ by taking $g_{2}=f_{2}=0$, there will be $15\%$ deviation
on the decay branching ratio.

\begin{table}[h]
\caption{Upper table is the branching ratio for $\Lambda _{b}\rightarrow
\Lambda \nu \bar{\nu}$ decay with different ratio $g_{i}/f_{i}$; lower one
shows the branching ratio while excluding $f_{2}$ and $g_{2}$.}
\begin{center}
\begin{tabular}{|c|c|c|c|c|c|c|c|}
\hline
$g_{1}/f_{1}=g_2/f_2$ & 0.80 & 0.90 & 0.95 & 1.00 & 1.10 & 1.15 & 1.20 \\
\hline
$10^{5}B(\Lambda _{b}\rightarrow \Lambda \nu \bar{\nu})$ & $1.498$ & $1.530$
& $1.547$ & $1.566$ & $1.606$ & $1.627$ & $1.650$ \\ \hline\hline
$g_{1}/f_{1}\ (g_2=f_2=0)$ & 0.80 & 0.90 & 0.95 & 1.00 & 1.10 & 1.15 & 1.20
\\ \hline
$10^{5}B(\Lambda _{b}\rightarrow \Lambda \nu \bar{\nu})$ & $1.584$ & $1.684$
& $1.739$ & $1.796$ & $1.920$ & $1.986$ & $2.055$ \\ \hline
\end{tabular}
\end{center}
\end{table}

In Figures 1 and 2, we show the differential decay branching ratio as
functions of the $\Lambda $ energy. We note that in Figure 1 there is a
turning point around $E_{\Lambda }\approx 1.9$ GeV. The ratios for $%
g_{i}/f_{i}>1$ are higher than that of $g_{i}/f_{i}\leq 1$ in the higher $%
E_{\Lambda }$ region, whereas it is opposite for the lower one. The reason
is due to the second term of $(|F_{1}|^{2}-|G_{1}| ^{2}) q^{2}$ in Eq. (\ref
{I1}). While lowering $E_{\Lambda }$, $q^{2}$ will be increased; and if $
% g_{i}/f_{i}<1$, the branching ratio will be decreased. As $%
E_{\Lambda }$ is over $1.9$ GeV, because of entering the small $q^{2}$
region, the term becomes less important.

\begin{table}[h]
\caption{The branching ratio for the different values of the Borel parameter
$M$.}
\begin{center}
\begin{tabular}{|c|c|c|c|}
\hline
$M$ & $1.5$ & $1.7$ & $1.9$ \\ \hline
$10^5B(\Lambda _{b}\rightarrow \Lambda \nu \bar{\nu})$ & $1.780$ & $1.566$ &
$1.554$ \\ \hline
\end{tabular}
\end{center}
\end{table}

For completeness, in Table 2 we present the decay branching ratios for
different values of the Borel parameter and in Figure 3 we show the
differential decay branching ratio with $\Lambda $ energy. We see that for
the smaller Borel parameter there is a larger deviation $(14\%) $ comparing
with that for $M=1.7$.

Finally, it is interesting to note that by defining
\begin{equation}
\bar{\alpha}_{B}=\frac{\int d\Gamma ^{0}\alpha _{B}dE_{\Lambda }}{\int
d\Gamma ^{0}dE_{\Lambda }},\ \ \ (B=\Lambda _{b},\Lambda) \,,  \label{apol}
\end{equation}
we have that $\bar{\alpha}_{\Lambda }\approx -1$ and $\bar{\alpha }_{\Lambda
_{b}}\approx -0.33$. \newline

\noindent $\bullet$ {\em New Physics}

If the hadronic sector involves only the left-handed interactions, from Eqs.
(\ref{I1}) and (\ref{ffgg}) we see that the decay rate depends on $\left|
C_{L}\right| ^{2}$. However, if the right-handed current interaction is
included, the dependence becomes $\left(|C_{R}| ^{2}+|C_{L}|
^{2}+Re\;C_{L}C_{R}^{*}\right)$. Since the interference term is associated
with a large product of form factors, $f_{1}g_{1}$, even for a small $C_{R}$
case, the physics beyond the standard model
 still makes a sizable effect.
In Table 3, we take a few allowed sets of $(\delta_{L},\delta_{R})$
from new physics and show the decay branching ratio of $\Lambda
_{b}\rightarrow \Lambda \nu \bar{\nu}$.

\begin{table}[h]
\caption{ The branching ratio from new physics for the different parameters
of $\delta _{L}$ and $\delta_{R}$ with $g_{i}/f_{i}=1$.}
\begin{center}
{\small
\begin{tabular}{|c|c|c|c|c|}
\hline $(\delta _{L},\ \delta_{R}$) & $(-2.25, -0.50)$ & $(-1.25,
-1.50)$ & $(0.10, 0.25)$ & $(0.50, 2.25)$ \\ \hline
$10^5B(\Lambda _{b}\rightarrow \Lambda \nu \bar{\nu})$ & $4.32$ & $4.51$ & $%
2.65$ & $19.5$ \\ \hline
\end{tabular}
}
\end{center}
\end{table}

\subsection{Polarization asymmetries}

To discuss the numerical values of the $\Lambda$ polarizations we assume
that $I_2/I_1\hat{e}_L\cdot \hat{\xi}_{\Lambda_b}$ is small so that we shall
neglect this term in our calculations. We also assume that $\hat{\xi}%
_{\Lambda }=0$ when we study the $\Lambda_b$ polarization. For the form
factors of $f_{i}$ and $g_{i}$, we will use the relations in Eq. (\ref{hqet}%
) and consider the cases with and without $f_{2}$ and $g_{2}$. To illustrate
the numerical values of the polarizations, we define the average
polarization asymmetries as
\begin{equation}
\bar{P}_{i}=\int_{E_{\min }}^{E_{\max }}P_{i}dE_{\Lambda }/M_{\Lambda
_{b}}\,,\quad \quad i=\Lambda_b,L,N,T  \label{avpo}
\end{equation}
where $E_{\max }=\left( M_{\Lambda _{b}}^{2}+M_{\Lambda }^{2}\right) /\left(
2M_{\Lambda }\right) $ and $E_{\min }=M_{\Lambda }$.\newline

\noindent $\bullet$ {\em In the standard model}

As discussed before, if there exist only left-handed interactions, the
coupling dependence for the longitudinal and normal polarization asymmetries
will be cancelled between the numerator and denominator. While for the
transverse part, since there are no CP violating phase and long distance
effect, it is expected to be zero. It is clear that the theoretical
uncertainties for the polarizations are from the hadronic transition form
factors and the CKM matrix elements. %%%%%%%%%%%%%%%%%%%%

\begin{table}[h]
\caption{ The average polarization of $\Lambda_{b}$ and longitudinal and
normal polarizations of $\Lambda$ with and without $f_{2}$ and $g_{2}$ in
the standard model}
\begin{center}
\begin{tabular}{|c|c|c|c|}
\hline
&  &  &  \\
$10^2\bar{P}_{B_i}$ & $10^2\bar{P}_{\Lambda_{b}}$ & $10^2\bar{P}_{L}$ &
$10^2\bar{P}_{N} $ \\ \hline
$f_1/g_1=f_2/g_2=1$ & $-7.40$ & $-31.30$ & $5.42$ \\ \hline
$f_1/g_1=1,f_2=g_2=0$ & $-8.66$ & $-27.12$ & $0$ \\ \hline
\end{tabular}
\end{center}
\end{table}

In Table 4, we show the average longitudinal and normal polarization
asymmetries of $\Lambda$ and the polarization $\Lambda_b$ with and without $%
f_{2}$ and $g_{2}$ in the standard model. From the table, we see that the
effects of $f_{2}$ and $g_{2}$ are between $13-17\%$. In the HQET limit, the
dominant term of the $\Lambda$ normal polarization asymmetry is proportional
to $-( f_{1}f_{2}+g_{1}g_{2})q^{2}\,p_{\Lambda }\cdot q$ and this
contribution is negligible when $f_{2}$ and $g_{2}$ are small. \newline

\noindent $\bullet$ {\em New Physics}

To search for the new physics effects, in Figure 4, we plot the contour
diagrams with several fixed values of the decay branching ratio and the
longitudinal polarization of $\Lambda$.
Here we have assumed that there are no phases for $\delta_{L,R}$.
 We note that $|C_{R}| >|C_{L}|$
corresponds to $\alpha _{\Lambda }>0$, while $|C_{R}| <|C_{L}|$ is for $%
\alpha _{\Lambda }<0$, since $\alpha_{\Lambda} $ is related to $%
|C_{R}|^{2}-|C_{L}|^{2}$. The forbidden regions in second and
fourth quadrants denote not only $\alpha _{\Lambda }>1$ but also
$I_{2(3)}/I_{1}>1$. Therefore, we obtain a further constraint on
$\delta_{L,R}$ that $(1+\delta_{L})$ and $\delta_{R}$ should take
the same sign in order to have the condition of $\alpha _{\Lambda
}\leq 1$. When the decay branching ratio and $\alpha _{\Lambda }$
in $\Lambda _{b}\rightarrow \Lambda \nu \bar{\nu}$ are measured,
we can determine the magnitude of $C_{L}$ and $C_{R}$ and the
relative sign (same sign) of them but not the individual.

 From Eqs. (\ref{I5}) and (\ref{ffgg}), we see that if the theory involves
only the left-handed interaction, $i.e$, $\delta_{L}\neq 0$ and $\delta
_{R}= 0$, $P_N$ is the same as that in the standard model because of the
cancellation of the coupling constants. For the case where $\delta _{R}\neq
0 $, the dominant terms for $P_N$ are proportional to $%
f_{1}^{2}C_{L}C_{R}^{*}$ and $f_{1}f_{2} ( | C_{R}| ^{2}+| C_{L}| ^{2}) $.
As we know that $f_{2}<0$, $f_{2}\ll f_{1}$ and $C_{L}C_{R}^{*}>0$ from the
constraint of $\bar{\alpha }_{\Lambda }\leq 1$, even with a small value of $%
| C_{R}| $, the sign of $P_N$ can be changed from the positive (SM-like
model) to negative. If the opposite sign of $P_{ N}$ is measured
experimentally, it clearly tells us that there is new physics of the
right-handed interaction.
In Table 5, to show the new physics affect for the
polarization asymmetries of polarized $\Lambda _{b}$ and $\Lambda $ in the
di-neutrino decay,
we list the average $\Lambda $ longitudinal and normal and $\Lambda _{b}$
polarization asymmetries with the same sets of $(\delta_{L},\delta_{R})$
as Table 3.
The distributions for the
polarizations of $\Lambda_b$ and $\Lambda$ with respect to the $\Lambda $
energy are shown in Figures $5-6$, respectively.

\begin{table}[h]
\caption{ The average polarization asymmetries for different values of $%
\delta_{L}$ and $\delta _{R}$ from new physics with $g_{i}/f_{i}=1$.}
\begin{center}
\begin{tabular}{|c|c|c|c|c|}
\hline
& &  &  &  \\
$\delta_{L}$ & $\delta_R$
&
$10^2\bar{P}_{\Lambda_{b}}$ & $10^2\bar{P}_{L}$ & $10^2\bar{P}_{N}$
\\ \hline
$-2.25$ &$-0.50$ & $-4.91$ & $-14.68$ & $-8.59$ \\
$-1.25$ &$-1.50$ & $6.81$ & $23.05$ & $-2.84$ \\
$0.10$&$0.25$ & $-6.38$ & $-20.67$ & $-4.77$ \\
$0.50$&$2.25$ & $2.51$ & $7.11$ & $-11.42$ \\ \hline
\end{tabular}
\end{center}
\end{table}

For the transverse polarization asymmetry $(P_T)$ which is
related to CP violation, from Eq. (\ref{I6}) we see that it depends on $Im(
1+\delta _{L}) \delta _{R}^{*}$. Thus, a non-vanished CP violating phase $%
\delta _{L}$ or $\delta _{R}$ will induce $P_T$. In Table 6, we just show a
few possible sets of $\delta _{L}$ and $\delta _{R} $, where the blank
values denote the exclusion by the condition of the longitudinal
polarization to be less than $1$. From the table, we see that the CP
violating polarization can
be large. In Figure 8, we show $P_T$ in terms of the $\Lambda $ energy.

\begin{table}[h]
\caption{ The average transverse polarization asymmetry ($\bar{P}_T$) for CP
violating theories with different complex parameters of $\delta_{L,R}$ and $%
g_{i}/f_{i}=1$.}
\begin{center}
{\small
\begin{tabular}{|c|c|c|c|c|}
\hline
&  &  &  &  \\
$10^2\bar{P}_T$ & $\delta_{R}=0.1+i0.05$ & $\delta _{R}=1.5+i0.5$ & $%
\delta_{R}=-0.1-i0.05$ & $\delta_{R}=-1.5-i1.0$ \\ \hline
$\delta _{L}=0.1+i0.05$ & $1.39$ & $3.22$ & $--$ & $--$ \\
$\delta _{L}=0.5+i0.1$ & $1.00$ & $3.11$ & $--$ & $--$ \\
$\delta _{L}=-2.1+i0.5$ & $--$ & $--$ & $2.53$ & $1.50$ \\
$\delta _{L}=-2.5+i1.5$ & $--$ & $--$ & $1.88$ & $16.39$ \\ \hline
\end{tabular}
}
\end{center}
\end{table}

\section{Conclusions}

{\normalsize
We have studied the decay of $\Lambda_b\to \Lambda\nu \bar{\nu}$
with the polarized baryons of $\Lambda _{b}$ and $\Lambda $. The general
form for the decay branching rate and the polarizations of $\Lambda _{b}$
and $\Lambda$ in terms of the general hadronic form factors have been given.
}

{\normalsize In the standard model, we have found that the decay branching
ratio of $\Lambda_b\to \Lambda\nu\bar{\nu}$ is between $1.5$ to $2.0\times
10^{-5}$. The average longitudinal polarization of $\Lambda$ is around $30\%$
while that of the normal one is small. Moreover, since there is no CP
violating phase from $V_{tb}V_{ts}^*$, the $\Lambda$ transverse polarization
is expected to be zero. The magnitude of the average $\Lambda_b$
polarization is below $10\%$. }

{\normalsize With new physics, such as the possible right-handed interaction
of $\delta_R=0.50$, the decay branching ratio can be as large as $4\times
10^{-5}$, and the magnitude of $\bar{P}_L$ and $\bar{P}_{\Lambda_b}$ become
smaller while $\bar{P}_N$ gets larger. In the CP violating theories, the CP
violating transverse $\Lambda$ polarization can be up to $16\%$, which could
be accessible in future experiments. }

\vspace{.5cm}

%\newpage
\noindent {\bf Acknowledgments}

{\normalsize This work was supported in part by the National Science Council
of the Republic of China under contract numbers NSC-89-2112-M-007-054 and
NSC-89-2112-M-006-004. }

{\normalsize \newpage }

{\normalsize %\begin{references}
}

{\normalsize %\end{references}
}

{\normalsize \newpage %\section{Figures}
\begin{figcap}

\item
%Fig.1:
The differential decay branching ratio as a function of $\Lambda$ energy
with different values of $g_{i}/f_{i} \ (i=1,2)$.
The solid curve denotes
$g_{i}/f_{i}=1$.
The thick dashed, dash-dotted, and dotted curves stand for
$g_i/f_i=1.10$, $1.15$, and $1.20$, while the thin ones
for $g_i/f_i=0.95$, $0.90$, and $0.80$, respectively.

\item
%Fig.2:
 The differential decay branching ratio as a function of $\Lambda$ energy
with $f_2=g_2=0$ and different values $g_1/f_1$.
The legend of $g_1/f_1$ is the same as Figure 1.

\item
%Fig.3:
 The differential decay
branching ratio as a function of $\Lambda$ energy
with different values of the Borel parameter: $M=1.5$
(dashed line), $M=1.7$ (solid line),
$M=1.9$ (dot dashed line).

\item
%Fig.4:
The elliptic closed curves represent
the decay branching ratios from
inside in turn as (1.0, 1.3,
%1.566,
1.6, 2.5, 3.0)
$\times 10^{-5}$. The lines in first and third quadrants correspond to
$P_L=\alpha_{\Lambda }$ being
$\pm 1.0$,
$\pm 0.8$, $\pm 0.5$, and
$\pm 0.3$, respectively.

\item
%Fig.5:
 The distribution of $P_{\Lambda_{b}}$ as a function of
$E_{\Lambda}/M_{\Lambda_{b}}$
with various new physics parameters of $(\delta _L,\delta_{R})$,
where the solid, dotted, dashed, dense-dotted, and dash-dotted curves
correspond to $(\delta
_{L},0)$, $(0.10,0.25)$, $(-2.25.-0.50)$,  $(0.50,2.25)$, and
$(-1.25,-1.50)$, respectively, and
$\delta _{L}$ expresses arbitrary value.

\item
%Fig.6:
 The distribution of $P_L$ as a function of
$E_{\Lambda}/M_{\Lambda_{b}}$. Legend is the same as Figure 5.

\item
%Fig.7:
 The distribution of $P_N$ as a function of
$E_{\Lambda}/M_{\Lambda_{b}}$. Legend is the same as Figure 5.

\item
%Fig.8:
 The distribution of $P_T$ as a function of
$E_{\Lambda}/M_{\Lambda_{b}}$
with various new physics parameters of $(\delta _L,\delta_{R})$,
where the solid, dotted, dashed, and dash-dotted curves
correspond to
$( -2.1+0.5i,-0.1-0.05i) $, $(-2.5+1.5i,-1.5-i)$, $( -2.1+0.5i,-1.5-i)$,
and $(-2.5+1.5i,-0.1-0.05i)$, respectively.

\end{figcap}
}

{\normalsize \newpage
\begin{figure}[h]
{\normalsize \includegraphics{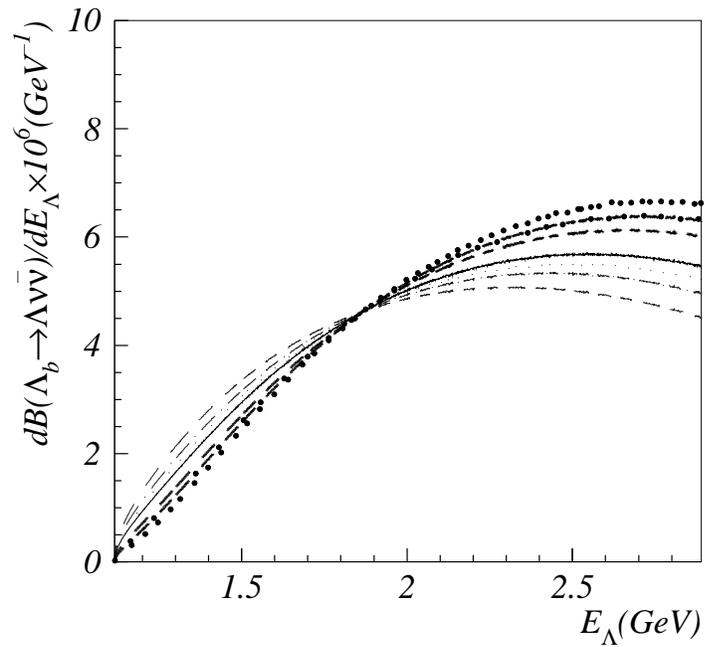} \vskip 11cm  }
\caption{ The differential decay branching ratio as a function of $\Lambda$
energy with different values of $g_{i}/f_{i} \ (i=1,2)$. The solid curve
denotes $g_{i}/f_{i}=1$. The thick dashed, dash-dotted, and dotted curves
stand for $g_i/f_i=1.10$, $1.15$, and $1.20$, while the thin ones for $%
g_i/f_i=0.95$, $0.90$, and $0.80$, respectively. }
\end{figure}
}

{\normalsize \newpage
\begin{figure}[h]
{\normalsize \includegraphics{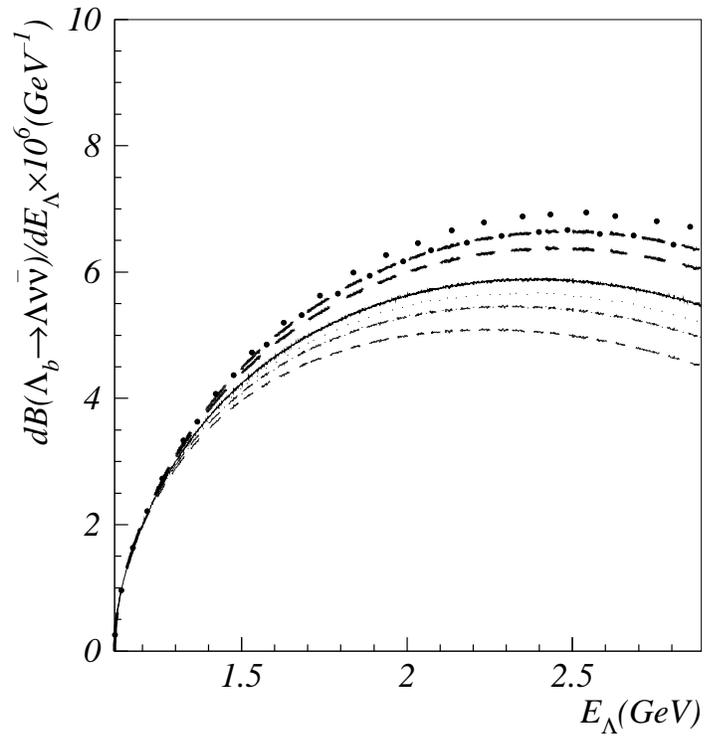} \vskip 11cm  }
\caption{ The differential decay branching ratio as a function of $\Lambda$
energy with $f_2=g_2=0$ and different values $g_1/f_1$. The legend of $%
g_1/f_1$ is the same as Figure 1. }
\end{figure}
}

{\normalsize \newpage
\begin{figure}[h]
{\normalsize \includegraphics{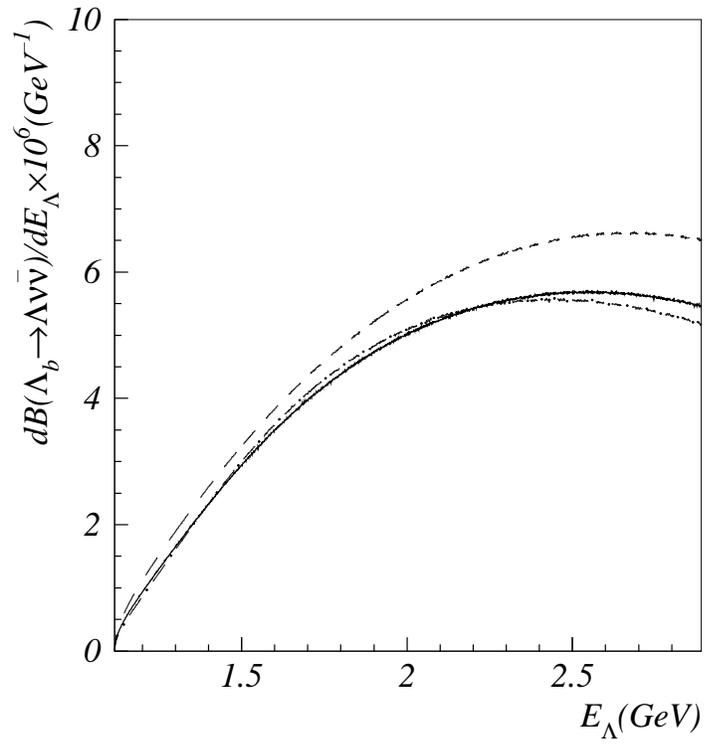} \vskip 11cm  }
\caption{ The differential decay branching ratio as a function of $\Lambda$
energy with different values of the Borel parameter: $M=1.5$ (dashed line), $%
M=1.7$ (solid line), $M=1.9$ (dot dashed line). }
\end{figure}
}

{\normalsize \newpage
\begin{figure}[h]
{\normalsize \includegraphics{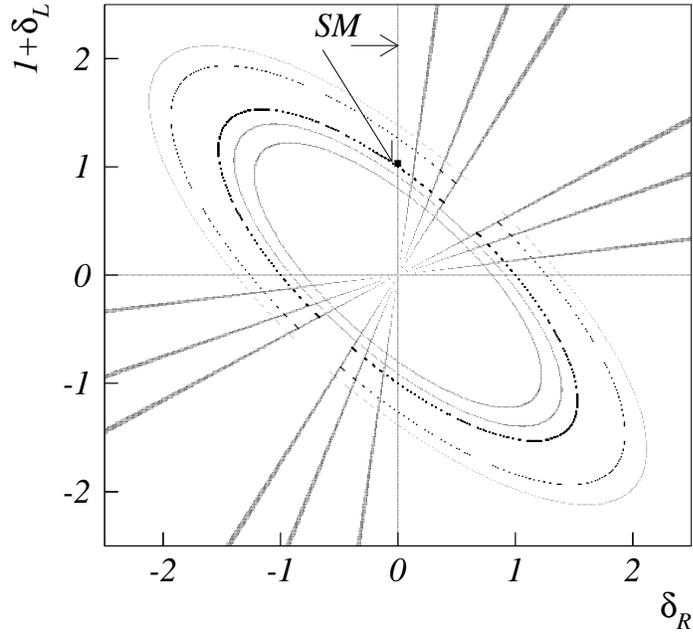} \vskip 11cm  }
\caption{ The elliptic closed curves represent the decay branching ratios
from inside in turn as (1.0, 1.3, 1.6, 2.5, 3.0) $\times 10^{-5}$. The lines
in first and third quadrants correspond to $P_L=\alpha_{\Lambda }$ being $%
\pm 1.0$, $\pm 0.8$, $\pm 0.5$, and $\pm 0.3$, respectively. }
\end{figure}
}

{\normalsize \newpage
\begin{figure}[h]
{\normalsize \includegraphics{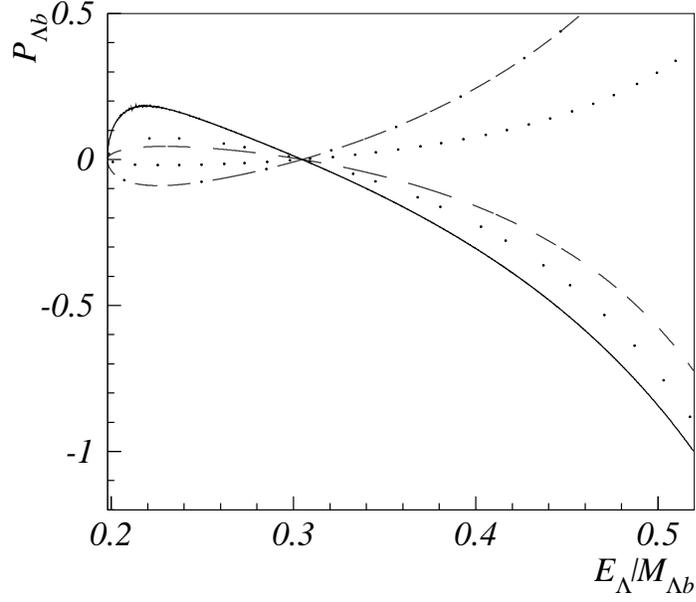} \vskip 11cm  }
\caption{ The distribution of $P_{\Lambda_{b}}$ as a function of $%
E_{\Lambda}/M_{\Lambda_{b}}$ with various new physics parameters of $(\delta
_L,\delta_{R})$, where the solid, dotted, dashed, dense-dotted, and
dash-dotted curves correspond to $(\delta _{L},0)$, $(0.10,0.25)$, $%
(-2.25.-0.50)$, $(0.50,2.25)$, and $(-1.25,-1.50)$, respectively, and $%
\delta _{L}$ expresses arbitrary value. }
\end{figure}
}

{\normalsize \newpage
\begin{figure}[h]
{\normalsize \includegraphics{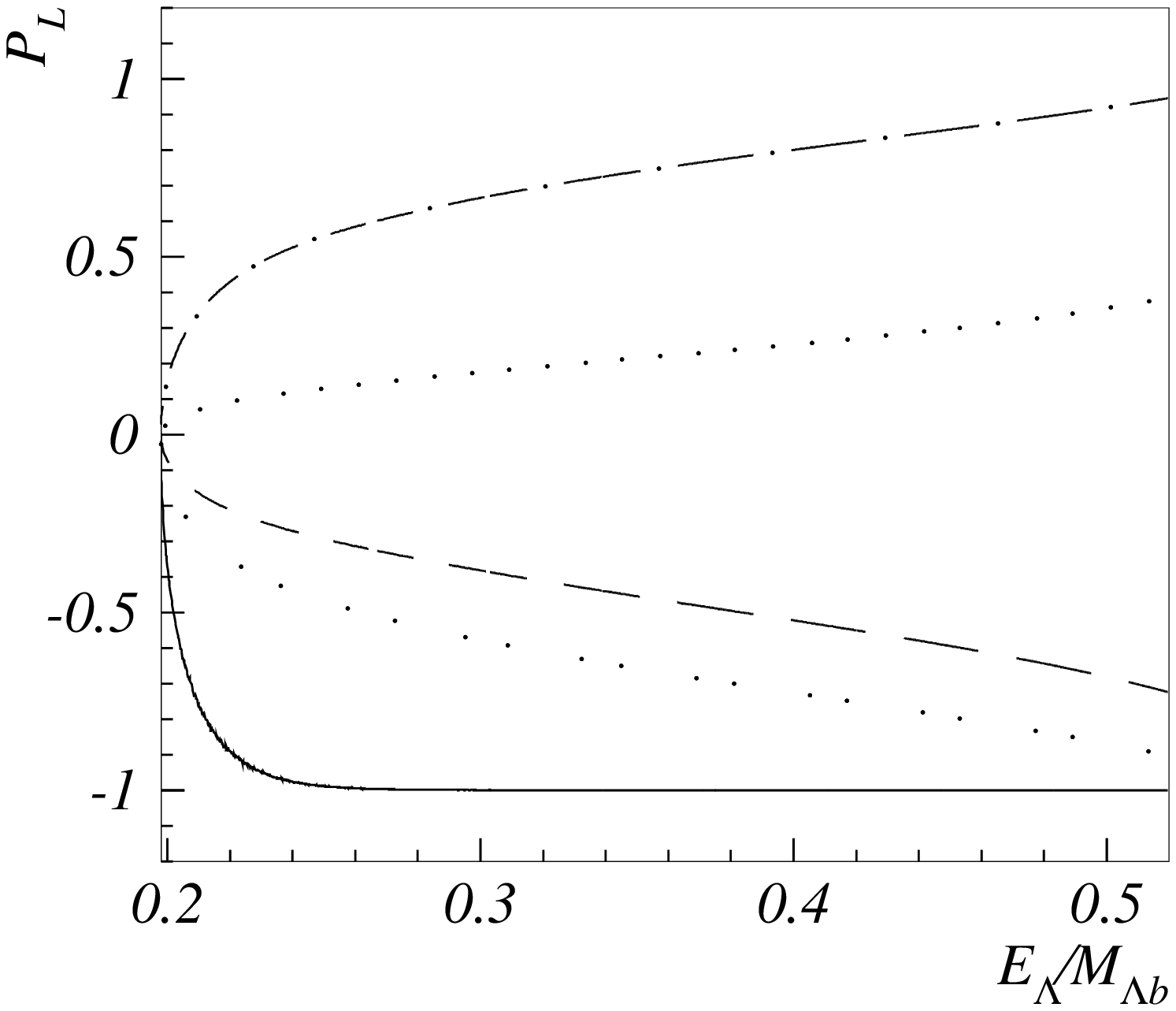} \vskip 11cm  }
\caption{ The distribution of $P_L$ as a function of $E_{\Lambda}/M_{%
\Lambda_{b}}$. Legend is the same as Figure 5. }
\end{figure}
}

{\normalsize \newpage
\begin{figure}[h]
{\normalsize \includegraphics{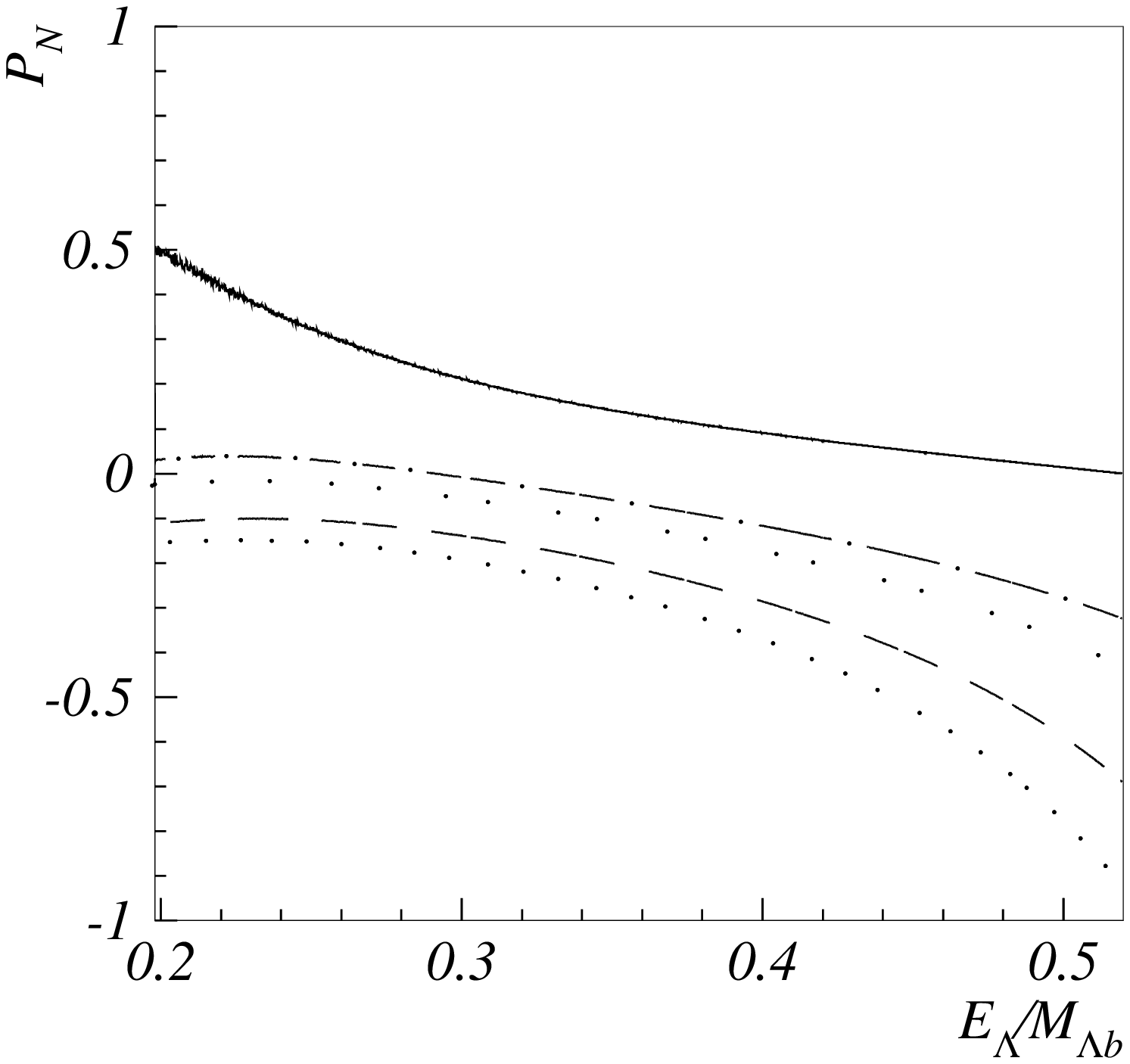} \vskip 11cm  }
\caption{ The distribution of $P_N$ as a function of $E_{\Lambda}/M_{%
\Lambda_{b}}$. Legend is the same as Figure 5. }
\end{figure}
%Fig.7:
}

{\normalsize \newpage
\begin{figure}[h]
{\normalsize \includegraphics{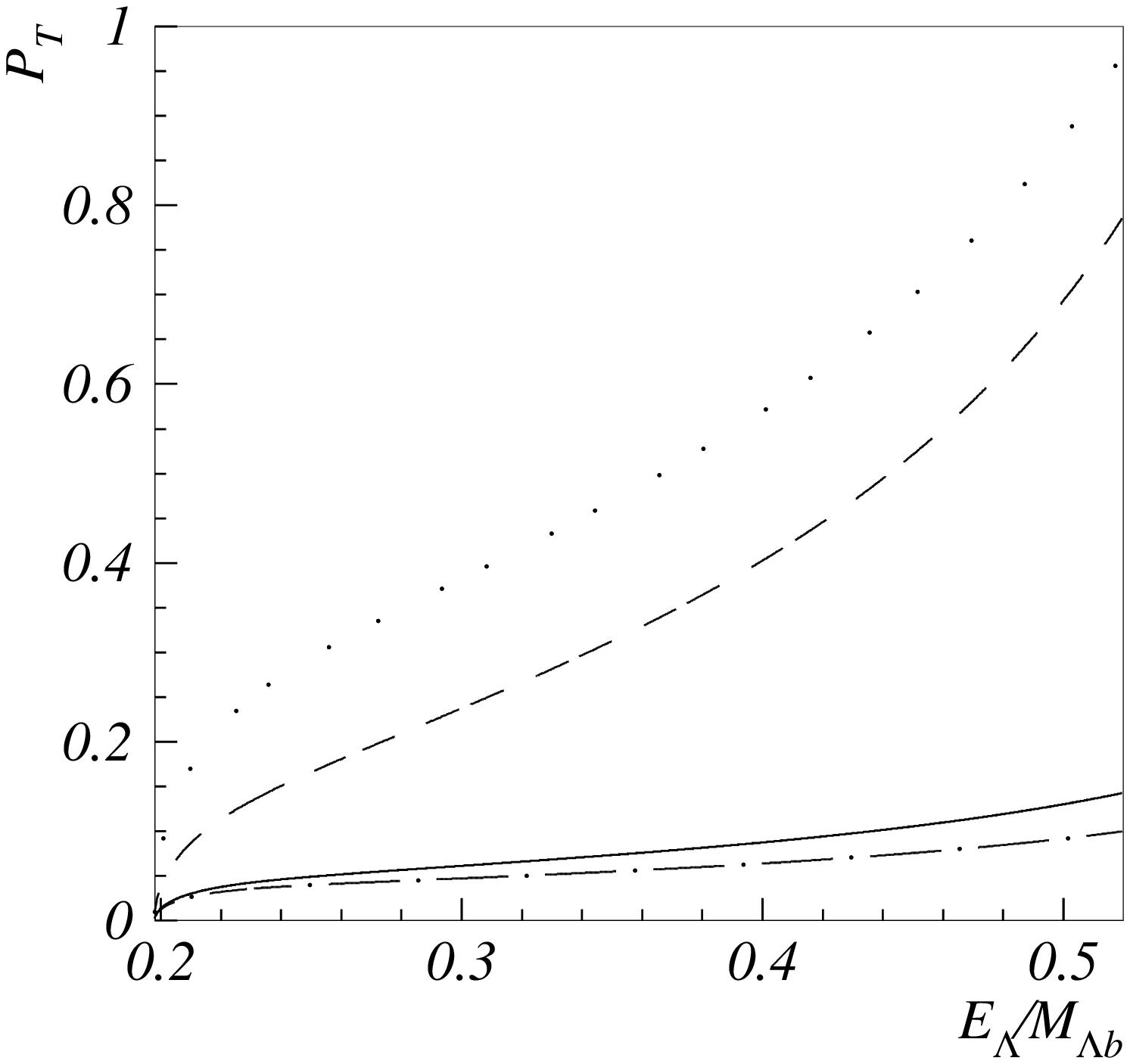} \vskip 11cm  }
\caption{ The distribution of $P_T$ as a function of $E_{\Lambda}/M_{%
\Lambda_{b}}$ with various new physics parameters of $(\delta _L,\delta_{R})$%
, where the solid, dotted, dashed, and dash-dotted curves correspond to $(
-2.1+0.5i,-0.1-0.05i) $, $(-2.5+1.5i,-1.5-i)$, $( -2.1+0.5i,-1.5-i)$, and $%
(-2.5+1.5i,-0.1-0.05i)$, respectively. }
\end{figure}
%Fig.8:
}


\begin{thebibliography}{99}
\bibitem{cleo}  {\normalsize CLEO Collaboration, M. S. Alam et. al., {\em %
Phys. Rev. Lett.} {\bf 74} (1995) 2885. }

\bibitem{ckm}  {\normalsize N. Cabibbo, {\em Phys. Rev. Lett.} {\bf 10}
(1963) 531; M. Kobayashi and T. Maskawa, {\em Prog. Theor. Phys.} {\bf 49}
(1973) 652. }

\bibitem{Lb1}  {\normalsize Chao-Shang Huang and Hua-Gang Yan, {\em Phys Rev.%
} {\bf D59} (1999) 114022. }

\bibitem{Lb2}  {\normalsize T.M. Aliev and M. Savci, {\em J. Phys.} {\bf G26}
(2000) 997. }

\bibitem{LD}  {\normalsize D. Rein and L. M. Sehgal, {\em Phys. Rev.} {\bf %
D39} (1989) 3325; J. Hagelin and L. S. Littenberg, {\em Prog. Part. Nucl.
Phys.} {\bf 23} (1989) 1; M. Lu and M. B. Wise, {\em Phys. Lett.} {\bf B324}
(1994) 461; C. Q. Geng, I. J. Hsu, and Y. C. Lin, {\em Phys. Lett.} {\bf B355%
} (1995) 569; C. Q. Geng, I. J. Hsu, and C. W. Wang, {\em Prog. Theor. Phys.}
{\bf 101} (1999) 937. }

\bibitem{Buchalla}  {\normalsize G. Buchalla and A.J. Buras, {\em Nucl. Phys.%
} {\bf B412} (1994) 106. }

\bibitem{Buras}  {\normalsize G. Buchalla, A. J. Buras and M. E.
Lautenbacher, {\bf Rev. Mod. Phys. 68} (1996) 1230. }

\bibitem{ALEPH}  ALEPH Collaboration, P. Perrodo {\it et al.,} in ICHEP '96,
Proceedings of the 28th International Conference on High Energy Physics,
Warsaw, Poland, edited by Z. Ajduk and A. Wroblewski (World
Scientific, Singapore, 1997).


\bibitem{MR}  {\normalsize T. Mannel, W. Roberts and Z. Ryzak {\em Nucl.
Phys.} {\bf B355} (1991) 38. }

\bibitem{MR2}  {\normalsize T. Mannel and S. Recksiegel, {\em J. Phys.} {\bf %
G24} (1998) 979. }
\end{thebibliography}
\end{document}